\documentclass{article}

\usepackage{arxiv}

\usepackage[utf8]{inputenc} 
\usepackage[T1]{fontenc}    
\usepackage{hyperref}       
\usepackage{url}            
\usepackage{booktabs}       
\usepackage{amsfonts}       
\usepackage{nicefrac}       
\usepackage{microtype}      
\usepackage{lipsum}		
\usepackage{graphicx}
\usepackage{natbib}
\usepackage{doi}
\usepackage{cite}
\usepackage{amsmath,amssymb,amsfonts}
\usepackage{algorithmic}
\usepackage{graphicx}
\usepackage{textcomp}
\usepackage{algorithm}
\usepackage{graphics}
\usepackage{epstopdf}
\usepackage{bm}
\usepackage{multicol}
\usepackage{multirow}
\usepackage{caption}
\usepackage{setspace}
\usepackage{listings}
\usepackage{lineno}
\usepackage{float} 
\usepackage{tabularx}
\newtheorem{lemma}{Lemma}
\usepackage{cuted}
\usepackage{lipsum}
\usepackage{makecell}
\usepackage{array}
\usepackage{booktabs}

\title{Dynamic probabilistic predictable feature analysis for multivariate temporal process monitoring}

\author{ \href{https://orcid.org/0000-0001-7054-2698}{\includegraphics[scale=0.06]{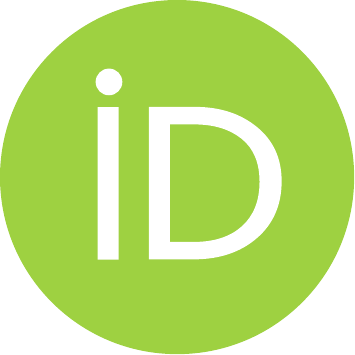}\hspace{1mm}Wei Fan} \\
	Key Laboratory of Energy Thermal Conversion\\ and Control of Ministry of Education\\
	Southeast University\\
	Nanjing 210096, China \\
	\texttt{wfan@seu.edu.cn} \\
	\And
	{\hspace{1mm}Qinqin Zhu}\footnotemark[1]\thanks{co-corresponding authors, equal contribution}\\
	Department of Chemical Engineering\\
	University of Waterloo\\
	ON N2L 3G1, Canada\\
	\texttt{qinqin.zhu@uwaterloo.ca} \\
	\And
{\hspace{1mm}Shaojun Ren}\footnotemark[1]\\
	Key Laboratory of Energy Thermal\\ Conversion and Control of Ministry of Education\\
	Southeast University\\
	Nanjing 210096, China \\
	\texttt{rsj@seu.edu.cn} \\
	\And
{\hspace{1mm}Liang Zhang} \\
	School of Instrument Science and Engineering\\ Key Laboratory of Micro-Inertial Instrument\\ and Advanced Navigation Technology\\ of Ministry of Education\\
	Southeast University\\
	Nanjing 210096, China \\
	\texttt{zhangliang418@seu.edu.cn} \\
	\And
{\hspace{1mm}Fengqi Si} \\
	Key Laboratory of Energy Thermal Conversion\\ and Control of Ministry of Education\\
	Southeast University\\
	Nanjing 210096, China \\
	\texttt{fqsi@seu.edu.cn} \\
}

\begin{document}
\maketitle

\begin{abstract}
Dynamic statistical process monitoring methods have been widely studied and applied in modern industrial processes. These methods aim to extract the most predictable temporal information and develop the corresponding dynamic monitoring schemes. However, measurement noise is widespread in real-world industrial processes, and ignoring its effect will lead to sub-optimal modeling and monitoring performance. In this article, a probabilistic predictable feature analysis (PPFA) is proposed for multivariate time series modeling, and a multi-step dynamic predictive monitoring scheme is developed. The model parameters are estimated with an efficient expectation-maximum algorithm, where the genetic algorithm and Kalman filter are designed and incorporated. Further, a novel dynamic statistical monitoring index, Dynamic Index, is proposed as an important supplement of $\text{T}^2$ and  $\text{SPE}$ to detect dynamic anomalies. The effectiveness of the proposed algorithm is demonstrated via its application on the three-phase flow facility and a medium speed coal mill.
\end{abstract}


\section{Introduction}
In the currentera of big data, industrial processes are equipped with a large number of sensors to measure different process variables. At the same time, massive amounts of historical data are collected and stored. On this basis, data-driven process monitoring has become a popular research topic due to its reliable performance and easy-to-implement characteristics \citep{severson2016perspectives,ge2017review,zhou2016autoregressive}. Multivariate statistical process monitoring method is a representative kind, including  principal component analysis (PCA) \citep{alcala2009reconstruction}, partial least squares \citep{li2010geometric} and canonical component analysis \citep{zhu2017concurrent}. As a dimensionality reduction algorithm, PCA decomposes the measurement space into principal component subspace and residual subspace, and realizes industrial process monitoring by designing relevant statistical indices. However, traditional PCA does not take temporal information into account, and thus tends to obtain sub-optimal performance.

To tackle the inevitable dynamics in the data samples, several extensions have been proposed. Ku et al. \citep{ku1995disturbance} proposed a dynamic PCA (DPCA), which employs augmented measurements with time lags and tries to explore the serial correlations between current and previous observations.
With the derived relations, statistical indices such as $\mathrm{T}^2$ and SPE are employed to monitor abnormal condition of industrial processes. Rato and Reis \citep{rato2013advantage} and Vanhatalo et al. \citep{vanhatalo2017structure} further improved DPCA in auto-correlation extraction and the selection of time lags respectively. Though some dynamics are exploited by these models, their internal structure is still static. Besides, DPCA fails to provide an explicit expression between latent variables and observed measurements, and the interpretability of the established model is limited \citep{dong2018novel}.

To address the aforementioned issues, Li et al. \citep{li2014new} proposed the vector auto-regressive (VAR) model, in which the concept of inner model was put forward. This method gives an explicit expression of the dynamic relation of latent variables, but its inner model is not consistent with the outer model, leading to sub-optimal monitoring performance \citep{guo2020multi}. Motivated by the concept of VAR, Dong and Qin \citep{dong2018novel} designed VAR in both inner and outer models, and developed a novel dynamic inner PCA (DiPCA) method to capture the most dynamic variations from time series data. Similarly, Richthofer and Wiskott \citep{richthofer2015predictable} proposed the predictable feature analysis (PFA) to extract dynamic latent variables that are as predictable as possible. Both DiPCA and PFA are efficient auto-regressive models, and their difference lies in the design of their objective functions. DiPCA builds a model by maximizing the covariance between the actual value and estimated value of the latent variable, while PFA aims to minimize the auto-regressive prediction error of the latent variable. 

In practical engineering applications, the variables are inevitable to be polluted by random noise, which, however, are not considered in the aforementioned methods. To provide the complete distribution of the data, the dynamic characteristics of process variables should be extracted through statistical patterns rather than deterministic manner \citep{zheng2016probabilistic}. Based on the above analysis, this article is dedicated to exploring the integration of probabilistic models and traditional VAR-based methods, and applying them for dynamic predictive process monitoring. Inspired by the state space expressions designed in probabilistic PCA (PPCA) \citep{tipping1999probabilistic} and probabilistic slow feature analysis (PSFA) \citep{guo2016monitoring}, we extend DiPCA and PFA to a probabilistic structure, referred to as probabilistic predictable feature analysis (PPFA). A high-order linear Markov state-space form is designed in PPFA to represent the weighted relations among its latent variables, and thus to model their dynamics. It is worthwhile to point out that PPFA has intrinsic advantages over DiPCA and PFA. First, deterministic methods fail to grasp the distribution of measurement noise, while PPFA overcomes this problem with a fully probabilistic framework.  On the other hand, inspired by the nonstationary PSFA \citep{scott2020holistic}, a monitoring statistic, Dynamic Index (DI), is derived to demonstrate the dynamic changes of studied systems, thereby providing a reliable guidance for improving control performance. The Expectation-Maximization (EM) algorithm \citep{dempster1977maximum} and Kalman filter \citep{bishop2006pattern} are adopted to estimate the model parameters of PPFA. The main contributions of this work are
\begin{itemize}
  \item A novel probabilistic extension of DiPCA and PFA, termed as PPFA,  is designed, which provides a full interpretation of the dynamic characteristics of both measurements and latent variables.  
  \item Multiple time lags are designed in PPFA to capture the actual dynamics involved in the data. 
  \item In M-step of the EM algorithm, the genetic algorithm (GA) \citep{goldberg1988genetic} is employed to optimize the weight coefficients of latent variables, which are difficult to be solved analytically.
  \item During the procedure of E-step, an expansion method of latent structure is proposed to estimate relevant expectations of the high-order dynamic system with Kalman filter.
  \item Based on the proposed PPFA model, three monitoring statistics, $\mathrm{T}^2$, SPE and DI, are developed for process monitoring.
\end{itemize}

The rest of this paper is organized as follows. Section 2 presents a brief introduction of DiPCA and PFA. In Section 3, the detailed derivation of PPFA as well as the novel PPFA-based dynamic process monitoring framework are demonstrated. Then, experiments on the three-phase flow facility and a medium speed coal mill are presented to testify the effectiveness of the proposed PPFA-based modeling and monitoring method. Finally, conclusions are drawn in Section 5.

\section{Preliminary}
\subsection{Dynamic inner Principal Component Analysis}
DiPCA is a dynamic extension of PCA that exploits temporal information from the data space to form its dynamic latent variables \citep{dong2018novel}. DiPCA aims to predict the current score with the past $s$ samples. Mathematically, it is expressed as
\begin{equation}\label{DiPCA_lv}
    t_k=\sum_{j=1}^s{\beta}_j t_{k-j}+e_k
\end{equation}
where $t_k=\mathbf{x}_k^\top \mathbf{w}$ is the latent score for the observed sample $\mathbf{x}_k$  at time $k$, $\mathbf{w}$ is the weight vector, $e_k$ is the modeling error, $s$ is the time lag, and ${\beta}_j$ is the auto-regressive coefficient. It is assumed that $e_k$ is white noise if $s$ is long enough, such that the estimated prediction of latent variable is expressed as
\begin{equation}\label{DiPCA_lv_model}
\begin{split}
    \widehat{t}_k &=\mathbf{x}_{k-1}^\top\mathbf{w}{\beta}_1+\cdots+\mathbf{x}_{k-s}^\top\mathbf{w}{\beta}_s\\
    &=\left[\mathbf{x}_{k-1}^\top\cdots\mathbf{x}_{k-s}^\top \right]\left(\bm{\beta}\otimes\mathbf{w}\right)
\end{split}
\end{equation}
where $\bm{\beta}=\left[{\beta}_1, {\beta}_2, \cdots, {\beta}_s\right]^\top$, and $\bm{\beta}\otimes\mathbf{w}$ is the Kronecker product.

Denote the data matrix as $\mathbf{X}=\left[\mathbf{x}_1^\top, \mathbf{x}_2^\top,  \cdots, \mathbf{x}_{n+s}^\top \right]^\top\in \mathbb{R}^{(n+s) \times m}$, and define a new data matrix containing temporal information $\mathbf{Z}_s=\left[\mathbf{X}_1, \mathbf{X}_2, \cdots, \mathbf{X}_{s} \right]$, where $\mathbf{X}_j=\left[\mathbf{x}_j^\top,  \mathbf{x}_{j+1}^\top,  \cdots,  \mathbf{x}_{n+j-1}^\top \right]^\top$, $j=1,2,\cdots,s$. Then the objective of DiPCA in Eq. \eqref{DiPCA_lv} can be represented as the following matrix form. 
 \begin{equation}\label{DiPCA_objective}
 \begin{split}
     &\max \limits_{\mathbf{w},\bm{\beta}} \ \ 
     \mathbf{w}^\top\mathbf{X}_{s+1}^\top\mathbf{Z}_s\left(\bm{\beta}\otimes\mathbf{w}\right)\\
     & \textup{s.t.}\ \ \|\mathbf{w}\|=1, \ \ \|\mathbf{\bm{\beta}}\|=1
\end{split}
\end{equation}
More details of extracting dynamic components for DiPCA can be found in ref. \citep{dong2018novel}.

\subsection{Predictable Feature Analysis}
Similar to DiPCA, PFA is also an auto-regressive model, and it focuses on measuring the predictability of the extracted latent variables \citep{richthofer2015predictable}. PFA defines that a good predictability is achieved when the latent score can be well predicted by a linear combination of $s$ past values. Given an $m$-dimensional temporal measurement $\mathbf{x}_k$, PFA aims to find an orthogonal transformation $\mathbf{W} \in \mathbb{R}^{m \times r}$, such that the projection $\mathbf{t}_k=\mathbf{x}_k^\top \mathbf{W}$ obtains the highest predictability. Mathematically, given the coefficient matrices  $\mathbf{B}_i\in\mathbb{R}^{r\times r}, 1\leq i \leq s$,  the prediction of $\mathbf{t}_k$ is 
 \begin{equation}\label{PFA_problem}
    \widehat{\mathbf{t}}_k=\sum_{i=j}^s \mathbf{B}_j\mathbf{t}_{k-j}=\sum_{j=1}^s \mathbf{B}_j\mathbf{W}^\top\mathbf{x}_{k-j}
 \end{equation}
where the modeling error is omitted. It is observed that both Eq. \eqref{DiPCA_lv} and Eq. \eqref{PFA_problem} share similar structure, and both of them aim to extract the most predictable latent variables. Their main difference is that the goal of DiPCA is to maximize the covariance between $t_k$ and $\widehat{t}_k$, while PFA solves the auto-regressive issue by minimizing the prediction error, which is expressed as
 \begin{equation}\label{PFA_objective}
     \min \limits_{\mathbf{W}} \ \ \langle\|\mathbf{W}^\top\mathbf{x}_k-\sum_{j=1}^s\mathbf{B}_j\mathbf{W}^\top\mathbf{x}_{k-j}\|^2\rangle
 \end{equation}
where $\langle\cdot\rangle$ means the average of signals over time. Readers can refer to ref. \citep{richthofer2015predictable} for the detailed PFA algorithm.

\section{Methodology}
\subsection{Probabilistic Predictable Feature Analysis}
In real-world industrial processes, the variables are inevitable to be polluted by random noise, which are not considered in both DiPCA and PFA. To improve the prediction performance, the dynamic characteristics of process variables should be extracted through statistical patterns rather than deterministic manner\citep{zheng2016probabilistic}. Therefore, this work is to extend DiPCA and PFA to a probabilistic structure with a state-space form, referred to as probabilistic predictable feature analysis (PPFA), which describes the process dynamics in a more compact and clearer manner. 

Given the collected samples $\mathbf{X}=\left[\mathbf{x}_1^\top \mathbf{x}_2^\top \cdots \mathbf{x}_{n+s}^\top \right]^\top$,  PPFA takes the following generative state-space form
\begin{equation}\label{PPFA_system}
\begin{split}
&\left\{
             \begin{array}{lr}
             \mathbf{t}_k =\sum_{j=1}^s \mathbf{B}_j\mathbf{t}_{k-j}+\mathbf{e}_k &\\
             \mathbf{x}_k =\mathbf{H}\mathbf{t}_k+ \bm{\varepsilon}_k &
             \end{array}
\right.\\
& \textup{s.t.}\ \ \mathbb{E} \left[\mathbf{t}_k\right]=\bm{0}, \ \ \mathbb{E} \left[\mathbf{t}_k \mathbf{t}_k^\top\right]=\mathbf{I}_{r}
\end{split}
\end{equation}
where $\mathbf{H}\in \mathbb{R}^{m \times r}$ represents the emission matrix, $\mathbf{B}_j=\text{diag}\left\{\beta_j^1,\cdots,\beta_j^r \right\}$, $(1\leq j \leq s)$ denotes the transition matrix, $\mathbf{e}_k \sim \mathcal{N}\left(\bm{0}, \bm{\Gamma}\right)$ with $\bm{\Gamma}=\textup{diag}\left\{\tau_1^2,\cdots,\tau_r^2\right\}$ is the Gaussian distributed noise, $\bm{\varepsilon}_k \sim \mathcal{N}\left(\bm{0}, \bm{\Sigma}\right)$ with $\bm{\Sigma}=\textup{diag}\left\{\sigma_1^2,\cdots,\sigma_m^2\right\}$ is the measurement noise, and $\mathbf{I}_{r}$ is an $r\times r$ identity matrix.

\begin{lemma}\label{Lemma1}
Given the constraints in Eq. \eqref{PPFA_system} that $\mathbb{E} \left(\mathbf{t}_k\right)=\bm{0}$ and $\mathbb{E} \left(\mathbf{t}_k \mathbf{t}_k^\top\right)=\mathbf{I}_{r}$, the relation between $\mathbf{B}_j$ and $\bm{\Gamma}$ is 
\begin{equation}\label{eq:Lemma1}
     \tau_i^2=1-\sum_{j=1}^s \beta_j^i\gamma_j^i \geq 0, \ \ 1\leq i \leq r
\end{equation}
where $\gamma_j^i=\mathrm{cov}\left(t_k^i,t_{k-j}^i\right)=\mathrm{cov}\left(t_{k+j}^i,t_k^i\right)$ is the autocovariance, and $t_k^i$ with $1\leq i \leq r$ represents the $i^\text{th}$ latent variable at time $k$.
\end{lemma}

The proof of \textit{Lemma \ref{Lemma1}}  is given in APPENDIX \ref{appendixA} , and the expression of $\gamma_j^i$ is also provided in the appendix.. For PPFA, the parameters needed to be estimated are summarized as $\bm{\Theta}=\left\{\mathbf{B}_j\ \ (1\leq j \leq s), \ \ \mathbf{H},\ \ \bm{\Gamma},\ \ \bm{\Sigma}\right\}$. 

The maximum-likelihood method is widely used for parameter optimization of probabilistic models. Given the temporal series $\mathbf{X}$, the complete data log-likelihood of the dynamic system in Eq. \eqref{PPFA_system} can be given by
\begin{equation}\label{eq:Lfunction}
    \begin{split}
         L\left(\bm{\Theta}\right)&=\log  p\left(\mathbf{X},\mathbf{T}|\bm{\Theta}\right) \\
     &=\log \prod_{k=s+1}^{n+s}\left[p\left(\mathbf{t}_k|\left\{ \mathbf{t}_{k-j} \right\}_{j=1}^s \right)p\left(\mathbf{x}_k| \mathbf{t}_k \right)\right]\\
     &\qquad \times \prod_{k=1}^{s}\left[p\left(\mathbf{t}_k\right)p\left(\mathbf{x}_k|\mathbf{t}_k\right)\right] \\
     &=\sum_{k=1}^s\log p\left(\mathbf{t}_k\right)+\sum_{k=1}^{n+s}\log p\left(\mathbf{x}_k| \mathbf{t}_k \right)\\ &\qquad +\sum_{k=s+1}^{n+s}p\left(\mathbf{t}_k|\left\{ \mathbf{t}_{k-j} \right\}_{j=1}^s \right)
    \end{split}
\end{equation}
where $\mathbf{T}=\left[\mathbf{t}_1^\top, \mathbf{t}_2^\top, \cdots, \mathbf{t}_{s+n}^\top\right]^\top\in \mathbb{R}^{(n+s)\times r}$ denotes the latent variables matrix. 

For each sample $\mathbf{x}_k$, based on the property of conditional independence, the corresponding high-order linear Markov Gaussian dynamic system $p\left(\mathbf{t}_k|\left\{ \mathbf{t}_{k-j} \right\}_{j=1}^s \right)$ in Eq. \eqref{eq:Lfunction} is expressed as
\begin{align}
    \label{eq:ARS1}
   p\left(\mathbf{t}_k|\left\{ \mathbf{t}_{k-j} \right\}_{j=1}^s \right) &\equiv p\left(\mathbf{t}_k|\left\{\mathbf{t}_{k-j}\right\}_{j=1}^s,\left\{\mathbf{B}_j\right\}_{j=1}^s,\bm{\Gamma}\right)\nonumber \\
   &=\prod_{i=1}^rp\left(t_k^i|\left\{t_{k-j}^i\right\}_{j=1}^s, \left\{\beta_{j}^i\right\}_{j=1}^s,\tau_i^2 \right)  
\end{align}
where
\begin{align}
   \label{eq:ARS2}
   &p\left(t_k^i|\left\{t_{k-j}^i\right\}_{j=1}^s, \left\{\beta_{j}^i\right\}_{j=1}^s,\tau_i^2 \right)=\mathcal{N}(\sum_{j=1}^s\beta_{j}^it_{k-j}^i, \tau_i^2)\\
   \label{eq:ARS3}
   &p\left(\mathbf{t}_1\right)=p\left(\mathbf{t}_2\right)=\cdots=p\left(\mathbf{t}_s\right)=\mathcal{N}(\bm{0},\mathbf{I}_r)
\end{align}

The probabilistic distribution of the mapping from $\mathbf{t}_k$ to $\mathbf{x}_k$ in Eq. \eqref{eq:Lfunction} is 
\begin{equation}\label{eq:ARS4}
     p\left(\mathbf{x}_k| \mathbf{t}_k \right) \equiv p\left(\mathbf{x}_k|\mathbf{t}_k, \mathbf{H}, \bm{\Sigma}\right)=\mathcal{N}\left(\mathbf{H}\mathbf{t}_k, \bm{\Sigma}\right)
\end{equation}

The complete specification of PPFA is defined in Eqs. \eqref{eq:Lfunction} to \eqref{eq:ARS4}, and the parameter set $\mathbf{\Theta}$ can be obtained via maximizing $L\left(\bm{\Theta}\right)$ in Eq. \eqref{eq:Lfunction}. It is noted that both stochastic design and multi-step time lags are considered in PPFA in Eq. \eqref{PPFA_system}, and several challenging issues should be addressed. Firstly, much more parameters in $\mathbf{\Theta}$ need to be determined, and a more efficient EM method should be designed. Further, the analytical solutions cannot be derived for $\beta_{j}^i$ in $\mathbf{B}_j$, and an optimization method should be integrated with the EM method. Additionally, for the high-order state-space model in Eq. \eqref{PPFA_system}, the expectations of the latent variables cannot be directly estimated with the traditional Kalman smoothing method \citep{bishop2006pattern}. In the following subsections, the traditional EM algorithm and Kalman smoothing method is optimized to address the aforementioned issues.

\subsection{Parameter Estimation Scheme}
The EM algorithm is a powerful method to solve the parameter estimation problem of probabilistic generative models through two steps of continuous iteration, namely E-step and M-step. In this section, we give detailed derivation process of EM algorithm applied in PPFA. The main body of parameter estimation process is the EM algorithm; however, it is difficult to obtain explicit analytical solutions for parameters $\beta_{j}^i$. Thus, GA algorithm is embedded into the EM process to construct the complete parameter optimization strategy.

\subsubsection{EM Algorithm}
Substituting the probability density function in Eqs. \eqref{eq:ARS1}-\eqref{eq:ARS4} into Eq. \eqref{eq:Lfunction}, we have
\begin{equation*}
    \begin{split}
    &L\left(\bm{\Theta}\right)    =-\frac{\left(n+s\right)\left(r+m\right)}{2}\log\left(2\pi\right)-\frac{1}{2}\sum_{k=1}^s\left(\mathbf{t}_k^\top\mathbf{t}_k\right)\\
    &\quad -n\sum_{i=1}^{r}\log(\tau_i)-\frac{1}{2}\sum_{k=1}^{n+s}\left[\left(\mathbf{x}_k-\mathbf{H}\mathbf{t}_k\right)^\top\Sigma^{-1}\left(\mathbf{x}_k-\mathbf{H}\mathbf{t}_k\right)\right]\\
    &\quad -\frac{\left(n+s\right)}{2}\log\left|\mathbf{\Sigma}\right|-\frac{1}{2}\sum_{k=s+1}^{n+s}\sum_{i=1}^{r}\frac{1}{\tau_i^2}\left[\tau_k^i-\sum_{j=1}^{s}\beta_j^i t_{k-j}^i\right]^2
    \end{split}
\end{equation*}

The $Q$-function is defined conditioned on the old parameter set $\bm{\Theta}^{\textup{old}}$ as
\begin{equation}\label{eq:Q}
    Q\left(\bm{\Theta},\bm{\Theta}^{\textup{old}}\right)=\mathbb{E}_{\mathbf{X},\bm{\Theta}^{\textup{old}}}\left\{ L\left(\bm{\Theta}\right)\right\}
\end{equation}
Then, in the M-step, new parameters are estimated through
\begin{equation}\label{eq:parm}
    \bm{\Theta}^{\textup{new}}=\mathop{\arg\max}_{\bm{\Theta}} \ \ Q\left(\bm{\Theta},\bm{\Theta}^{\textup{old}}\right)
\end{equation}

Since $\mathbf{B}_j$ $(1\leq j \leq s)$ and $\mathbf{\Gamma}$ are related to each other as shown in \textit{Lemma \ref{Lemma1}}, the parameter set to be estimated can be simplified to $\bm{\Theta}=\left\{\mathbf{B}_j,1\leq j \leq s, \mathbf{H},\bm{\Sigma}\right\}$. Then take derivatives of $Q$-function with respect to different parameters in $\bm{\Theta}$. First of all, for $\beta_j^i$ in $\mathbf{B}_j$, taking the derivation leads to
\begin{equation*}
\begin{split}
    \frac{\partial Q\left(\bm{\Theta},\bm{\Theta}^{\textup{old}}\right)}{\partial \beta_j^i}&=\frac{n\gamma_j^i}{2\left(1-\sum_{l=1}^{s}\beta_l^i\gamma_l^i\right)}-\frac{1}{2}\sum_{k=s+1}^{n+s}\frac{\gamma_j^i\left[ t_k^i-\sum_{l=1}^s\beta_l^it_{k-l}^i\right]^2}{\left(1-\sum_{l=1}^{s}\beta_l^i\gamma_l^i\right)^2}-\frac{1}{2}\sum_{k=s+1}^{n+s}\frac{2t_{k-j}^i\left[t_k^i-\sum_{l=1}^s\beta_l^it_{k-l}^i\right]}{1-\sum_{l=1}^{s}\beta_l^i\gamma_l^i} = 0
\end{split}
\end{equation*}
Then setting the above equation to zero, we have
\begin{equation}\label{eq:beta2}
\begin{split}
    &\left[{n\gamma_j^i-2\sum_{l=1}^{s}\beta_l^i\sum_{k=s+1}^{n+s}t_{k-j}^it_{k-l}^i } {+2\sum_{k=s+1}^{n+s}t_k^it_{k-j}^i}\right] \\
    & \times \left(1- \sum_{l=1}^{s}\beta_l^i\gamma_l^i\right) = 2\sum_{l=1}^{s}\sum_{g=l+1}^{s}\beta_l^i\beta_g^i\times \sum_{k=s+1}^{n+s}\left(t_{k-j}^it_{k-g}^i\right)\\
    & + \gamma_j^i\left[{\sum_{k=s+1}^{n+s}\left(t_k^i\right)^2-2\sum_{l=1}^s\beta_l^i\sum_{k=s+1}^{n+s}t_k^it_{k-l}^i} \right. \left.{+\sum_{l=1}^s\left(\beta_l^i\right)^2}\sum_{k=s+1}^{n+s}\left(t_{k-l}^i\right)^2\right]
\end{split}
\end{equation}

It is difficult to derive the analytical solution for Eq. \eqref{eq:beta2}. Alternatively, GA algorithm is employed to simplify the derivation, which is demonstrated in the next subsection.

For parameter matrix $\mathbf{H}$, differentiating $Q\left(\bm{\Theta},\bm{\Theta}^{\textup{old}}\right)$ with respect to it and setting it to zero result in
\begin{equation}\label{eq:Hnew}
   \mathbf{H}^{\text{new}}=\left(\sum_{k=1}^{n+s}\mathbf{x}_k\mathbf{t}_k^\top\right)\left(\sum_{k=1}^{n+s}\mathbf{t}_k\mathbf{t}_k^\top\right)^{-1}
\end{equation}

Similarly, the covariance parameter $\mathbf{\Sigma}$ is updated with 
\begin{equation}\label{eq:Sigma2}
   \mathbf{\Sigma}^{\text{new}}=\frac{1}{n+s}\sum_{k=1}^{n+s}\left(\mathbf{x}_k-\mathbf{H}^{\text{new}}\mathbf{t}_k\right)\left(\mathbf{x}_k-\mathbf{H}^{\text{new}}\mathbf{t}_k\right)^\top
\end{equation}

It is noted that $\mathbf{\Sigma}=\textup{diag}\left\{\sigma_1^2,\cdots,\sigma_m^2\right\}$, and each $\sigma_i^2$ in $\mathbf{\Sigma}$ can be calculated by
\begin{equation}\label{eq:Sigma3}
   \sigma_i^2=\frac{1}{n+s}\sum_{k=1}^{n+s}\left\{\left(x_k^i\right)^2-2\mathbf{h}_i^\top\mathbf{t}_kx_k^i+\mathbf{h}_i^\top\mathbf{t}_k\mathbf{t}_k^\top\mathbf{h}_i \right\}
\end{equation}
where $\mathbf{h}_i=[h_i^1,\cdots,h_i^r]^\top$ is the $i^\mathrm{th}$ row of $\mathbf{H}^{\text{new}}$.

During the process of M-step, the following expectations with respect to the latent variables $\mathbf{t}_k$ should be evaluated, which will be further utilized in the  E-step given the old parameter set $\bm{\Theta}^{\textup{old}}$.
\begin{align}
    &\mathbb{E}_{\mathbf{t}_k|\mathbf{x}_k,\bm{\Theta}^{\textup{old}}}\left[\mathbf{t}_k\right] \nonumber \\
    \label{eq:expectation}
    &\mathbb{E}_{\mathbf{t}_k|\mathbf{x}_k,\bm{\Theta}^{\textup{old}}}\left[\mathbf{t}_k\mathbf{t}_k^\top\right]\\
    &\mathbb{E}_{\mathbf{t}_k|\mathbf{x}_k,\bm{\Theta}^{\textup{old}}}\left[\mathbf{t}_k\mathbf{t}_{k-l}^\top\right] , \, 1\leqslant l \leqslant s \nonumber \\
    &\mathbb{E}_{\mathbf{t}_k|\mathbf{x}_k,\bm{\Theta}^{\textup{old}}}\left[\mathbf{t}_{k-l}\mathbf{t}_{k-g}^\top\right], \, 1\leqslant l, \, g \leqslant s\nonumber
\end{align}
Eq. \eqref{eq:expectation} will be further investigated in Subsection "Expectation Estimation Strategy".

\subsubsection{Genetic Algorithm}
The genetic algorithm is adopted to get the values of $\beta_j^i$ in Eq. \eqref{eq:beta2}. GA is an adaptive heuristic optimization algorithm proposed on the basis of natural selection in the theory of evolution \citep{goldberg1988genetic}. The main idea of GA is to model the system that satisfies the natural evolution conditions, and solve the optimization problem by randomly searching the pre-constructed search space. During the iterative process, the population is mutated and reorganized, and the fitness function is used to evaluate the reliability of individuals. In the end, individuals who adapt to the system environment are evolved, and the best one is selected as the parameter candidate \citep{garg2016hybrid}. 

To estimate $\beta_j^i$, the optimization model is constructed as follows. Based on Eq. \eqref{eq:beta2}, two reference indices $\mathrm{A}$ and $\mathrm{B}$ are defined as follows
\begin{align}
    \mathrm{A}&=\left[{n\gamma_j^i-2\sum_{l=1}^{s}\beta_l^i\sum_{k=s+1}^{n+s}t_{k-j}^it_{k-l}^i } {+2\sum_{k=s+1}^{n+s}t_k^it_{k-j}^i}\right] \nonumber\\
    \label{eq:beta2l}
    & \qquad \times \left(1-\sum_{l=1}^{s}\beta_l^i\gamma_l^i\right) \\
    \mathrm{B}&=2\sum_{l=1}^{s}\sum_{g=l+1}^{s}\beta_l^i\beta_g^i \sum_{k=s+1}^{n+s}\left(t_{k-j}^it_{k-g}^i\right) \nonumber \\
    & \qquad + \gamma_j^i\left[{\sum_{k=s+1}^{n+s}\left(t_k^i\right)^2-2\sum_{l=1}^s\beta_l^i\sum_{k=s+1}^{n+s}t_k^it_{k-l}^i} \right.\nonumber\\
    \label{eq:beta2r} 
    & \qquad \left.{+\sum_{l=1}^s\left(\beta_l^i\right)^2}\sum_{k=s+1}^{n+s}\left(t_{k-l}^i\right)^2\right]
\end{align}

Combined with \textit{Lemma \ref{Lemma1}}, the optimization problem for $\beta_j^i$ is formulated as
\begin{equation}\label{eq:GA1}
\begin{split}
    &\min_{\beta_j^i}\ \ f\left(\beta_j^i\right)=\left(A-B\right)^2\\
    &\textup{s.t.}\ \ 1-\sum_{j=1}^s \beta_j^i\gamma_j^i \geq 0
\end{split}
\end{equation}

To handle the inequality constraints, we further combine the Lagrange method to adjust the objective function as follows.
\begin{equation}\label{eq:GA2}
\min_{\beta_j^i}\ \ g\left(\beta_j^i\right)=
\left\{
             \begin{array}{lr}
             f\left(\beta_j^i\right),\ \ 1-\sum_{j=1}^s \beta_j^i\gamma_j^i \geq 0& \\
             f\left(\beta_j^i\right)-\lambda\left(1-\sum_{j=1}^s \beta_j^i\gamma_j^i\right), \text{ otherwise}&  
             \end{array}
\right.
\end{equation}
where $\lambda$ denotes a regularization coefficient which is greater than zero. The detailed modeling procedure of GA is provided in ref. \citep{goldberg1988genetic}. It is noted that GA is adopted as an illustrative instance to obtain the optimal parameters of PPFA, and other optimization algorithms such as gradient descent and particle swarm optimization \citep{maclaurin2015gradient, bergstra2012random,eggensperger2013towards,lorenzo2017particle} can also be designed. For large-scale processes, more efficient optimization methods such as gradient descent \citep{maclaurin2015gradient} is recommended.

\subsubsection{Expectation Estimation Strategy}
For a general first-order linear dynamic system, as proposed by Shang et al. \citep{shang2015probabilistic}, the expectations in Eq. \eqref{eq:expectation} can be easily estimated with Kalman smoothing \citep{bishop2006pattern}. Typically, the forward recursion and the backward recursion are employed to solve the problem, which, however, cannot be  directly used to estimate the desired expectations for the high-order system proposed in this work. 

To address this issue, we expand the dimension of the dynamic system designed in Eq. \eqref{PPFA_system} as follows, and adjust it to the general form of a  first-order linear dynamic system. Therefore, the general Kalman smoothing method can be easily adapted to perform the E-step.
\begin{equation}\label{eq:kal1}
\begin{split}
    &\mathbf{t}_k^s=\left[\mathbf{t}_k^\top,\ \ \mathbf{t}_{k-1}^\top, \ \ \cdots, \ \ \mathbf{t}_{k-s+1}^\top\right]^\top \in \mathbb{R}^{rs}, \\
    &\mathbf{e}_k^s=\left[\mathbf{e}_k^\top,\ \ \bm{0}, \ \ \cdots,\ \ \bm{0}\right]^\top\in \mathbb{R}^{rs}, \\
    &\bm{\Phi}_k=\left [
  \begin{matrix}
   \mathbf{B}_1, & \mathbf{B}_2, & \cdots, & \mathbf{B}_{s-1}, & \mathbf{B}_s \\
   \mathbf{I}, & \bm{0}, & \cdots, &  \bm{0},&  \bm{0} \\
   \vdots & \vdots & \ddots & \vdots& \vdots\\
   \bm{0}, & \bm{0}, & \cdots, &\mathbf{I},& \bm{0}
  \end{matrix} \right]\in \mathbb{R}^{rs\times rs},\\
  &\mathbf{H}_k=\left[\mathbf{H}, \ \ \bm{0}, \ \ \cdots, \ \ \bm{0}\right]\in \mathbb{R}^{m\times rs}
\end{split}
\end{equation}
\begin{equation}\label{eq:kal1a}
\begin{split}
  &\mathbf{\Gamma}_k=\left [
  \begin{matrix}
  \bm{\Gamma}, & \bm{0}, & \cdots, &  \bm{0},&  \bm{0} \\
  \bm{0}, & \bm{0}, & \cdots, &  \bm{0},&  \bm{0} \\
  \vdots & \vdots & \ddots & \vdots& \vdots\\
  \bm{0}, & \bm{0}, & \cdots, &  \bm{0},&  \bm{0} \\
  \end{matrix}\right]\in \mathbb{R}^{rs\times rs}
\end{split}
\end{equation}

Thereafter, the dynamic system proposed in Eq. \eqref{PPFA_system} is adapted into
\begin{equation}\label{PPFA2}
\left\{
             \begin{array}{lr}
             \mathbf{t}_k^s =\bm{\Phi}_k\mathbf{t}_{k-1}^s+\mathbf{e}_k^s &\\
             \mathbf{x}_k =\mathbf{H}_k\mathbf{t}_k^s+ \bm{\varepsilon}_k &\end{array}
\right.
\end{equation}
where the Gaussian noise $\mathbf{e}_k^s\sim \mathcal{N}\left(\bm{0}, \bm{\Gamma}_k\right)$. 

Correspondingly, the initial conditions given by Eq. \eqref{eq:ARS3} become
\begin{equation}\label{eq:initial_s}
    \mathbf{t}_s^s=\left[\mathbf{t}_s^\top\ \ \mathbf{t}_{s-1}^\top \ \ \cdots \ \ \mathbf{t}_{1}^\top\right]\sim \mathcal{N}\left(\bm{0}, \mathbf{I}_{rs}\right)
\end{equation}
where $\mathbf{I}_{rs}\in \mathbb{R}^{rs\times rs}$ is an identity matrix.

After the modification, the forward recursions of Kalman smoothing \citep{bishop2006pattern} are first applied to estimate the posterior distribution of $\mathbf{t}_k^s$ given the total past-time $\mathbf{x}_k$ and $\bm{\Theta}^\text{old}$, termed as $p\left(\mathbf{t}_k^s|\mathbf{x}_1,\cdots,\mathbf{x}_k,\bm{\Theta}^\text{old}\right) \sim \mathcal{N} \left( \bm{\mu}_k, \mathbf{V}_k \right) $, sequentially.
\begin{equation}\label{eq:fwd}
    \begin{split}
        &\bm{\mu}_k=\bm{\Phi}_k\bm{\mu}_{k-1}+\mathbf{K}_k\left(\mathbf{x}_k-\mathbf{H}_k\bm{\Phi}_k\bm{\mu}_{k-1}\right),\\
        &\mathbf{P}_{k-1}=\bm{\Phi}_k\mathbf{V}_{k-1}\bm{\Phi}_k^\top+\bm{\Gamma}_k,\\
        &\mathbf{V}_k=\left(\mathbf{I}-\mathbf{K}_k\mathbf{H}_k\right)\mathbf{P}_{k-1},\\
        &\mathbf{K}_k=\mathbf{P}_{k-1}\mathbf{H}_k^\top\left(\mathbf{H}_k\mathbf{P}_{k-1}\mathbf{H}_k^\top+\bm{\Sigma}\right)^{-1}
    \end{split}
\end{equation}
with the initialization
\begin{equation}\label{eq:fwd_ini}
    \begin{split}
        &\bm{\mu}_s=\mathbf{K}_1\mathbf{x}_s,\\
        &\mathbf{V}_s=\mathbf{I}-\mathbf{K}_1\mathbf{H}_k,\\
        &\mathbf{K}_s=\mathbf{H}_k^\top \left(\mathbf{H}_k\mathbf{H}_k^\top+\bm{\Sigma}\right)^{-1}
    \end{split}
\end{equation}
where $\bm{\mu}_k$ and $\mathbf{V}_{k}$ represent the mean vector and covariance matrix of the posterior distribution, respectively. $\mathbf{P}_k$ is the error covariance matrix of state estimate $\mathbf{t}_k^s$, and $\mathbf{K}_k$ denotes the Kalman gain matrix.

Afterward, utilizing the backward recursions \citep{bishop2006pattern} to further estimate the parameters of posterior distribution $p\left(\mathbf{T}|\mathbf{X},\bm{\Theta}^\text{old}\right)$.
\begin{equation}\label{eq:bwd}
    \begin{split}
        &\widehat{\bm{\mu}}_k=\bm{\mu}_k+\mathbf{J}_k\left(\widehat{\bm{\mu}}_{k+1}-\bm{\Phi}_k\bm{\mu}_k\right),\\
        &\widehat{\mathbf{V}}_k=\mathbf{V}_k+\mathbf{J}_k\left(\widehat{\mathbf{V}}_{k+1}-\mathbf{P}_k\right)\mathbf{J}_k^\top,\\
    \end{split}
\end{equation}
where $\mathbf{J}_k=\mathbf{V}_k\bm{\Phi}_k\mathbf{P}_k^{-1}$, and the initialization is expressed as
\begin{equation}\label{eq:bwd_ini}
    \widehat{\bm{\mu}}_{n+s}=\bm{\mu}_{n+s},\widehat{\mathbf{V}}_{n+s}=\mathbf{V}_{n+s}
\end{equation}

The derivation results between Eqs. \eqref{eq:kal1} and \eqref{eq:bwd_ini}, especially $\bm{\mu}_k$ and  $\mathbf{V}_k$, are important referent variables to evaluate the expectation terms in Eq. \eqref{eq:expectation}.

\begin{figure*}[h!]
    \centering
    \includegraphics[width=6in]{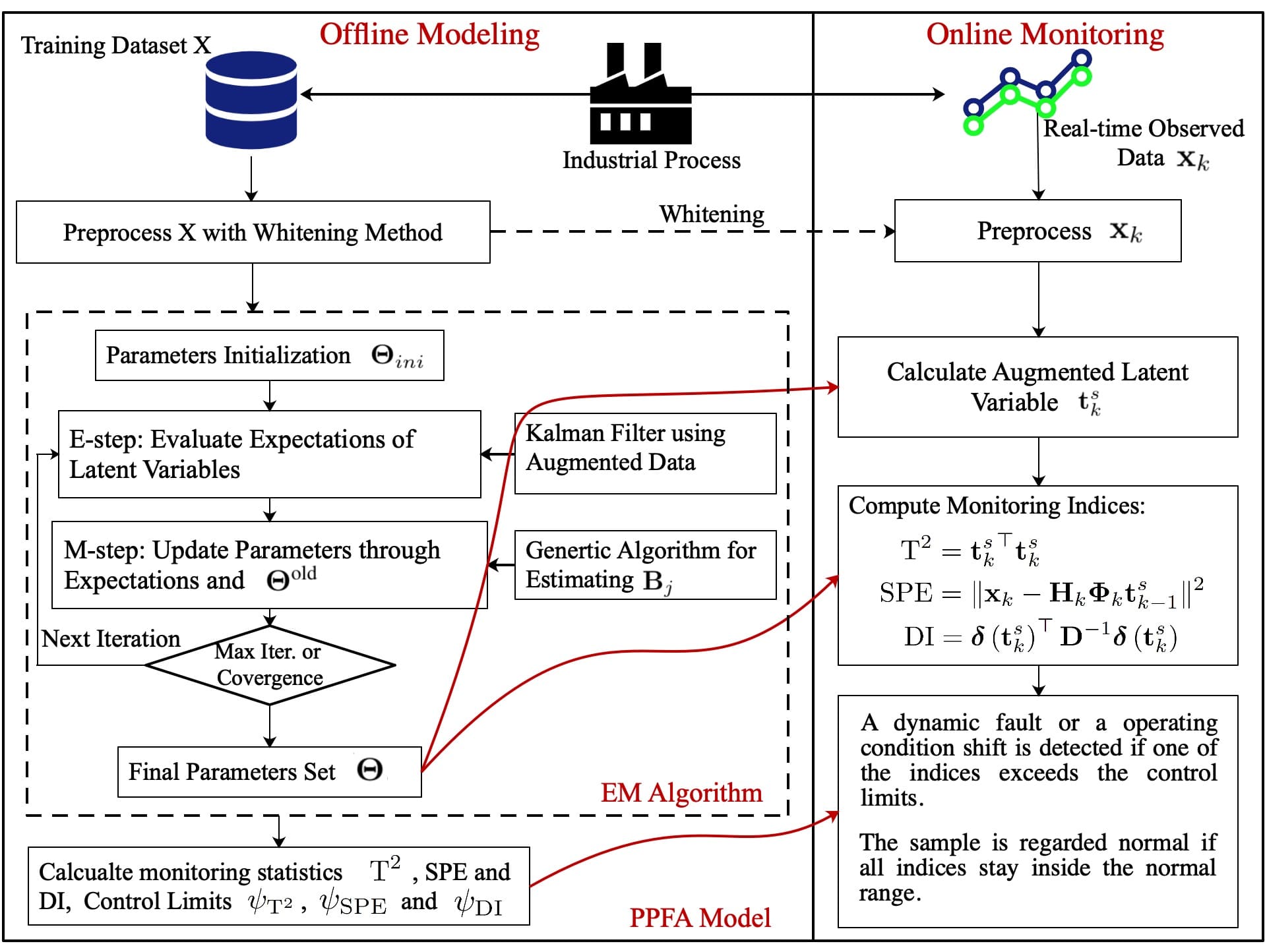}
    \caption{The schematic diagram of the proposed process monitoring method.}
    \label{pic:Recon}
\end{figure*}

\subsection{Parameter Determination}
In PPFA modeling, two parameters, the number of latent variables $r$ and the time lag $s$, need to be determined before executing the EM algorithm, and the hold-out method is adopted in this work. Firstly, the normal samples are separated into two subsets with 80\% as the training one and 20\% as the validation one, and deviations with various magnitudes are randomly added to different observed variables in the validation subset. A relatively wide range is initialized for $r$ and $s$ respectively, and for each pair of $r$ and $s$, the trained model is evaluated with two metrics, fault detection rate (FDR) and false alarm rate (FAR), which are defined as 
\begin{equation}\label{eq:FDR}
    FDR= \frac{TP}{TP+FN}
\end{equation}
\begin{equation}\label{eq:FAR}
    FDR= \frac{FP}{FP+TN}
\end{equation}
where TP and TN are the number of correctly detected abnormal and normal samples respectively, FN is the number of faults that are not detected, and FP is the number of normal samples which are regarded as anomalies incorrectly. The $\left(r,s\right)$ pair with the best monitoring performance in terms of FDR and FAR is chosen as the parameters of PPFA.

\subsection{Monitoring Statistics Design}
Once the PPFA model is established, monitoring statistical indices are of great significance when proceeding to the online monitoring part. For different indices, $\mathrm{T}^2$ and SPE are the most popular ones, which assume that the data follows a Gaussian distribution \citep{joe2003statistical}. With the parameters set $\bm{\Theta}$ obtained with EM and genetic algorithm, the augmented latent variable $\mathbf{t}_k^s=\bm{\Phi}_k\mathbf{t}_{k-1}^s+\mathbf{K}_k\left(\mathbf{x}_k-\mathbf{H}_k\bm{\Phi}_k\mathbf{t}_{k-1}^s\right)$ can be applied to define statistical indices.
\begin{align}\label{eq:T2}
    \mathrm{T}^2&={\mathbf{t}_k^s}^\top \mathbf{t}_k^s\\
    \label{eq:SPE}
    \mathrm{SPE}&=\|\mathbf{x}_k-\mathbf{H}_k\bm{\Phi}_k\mathbf{t}_{k-1}^s\|^2
\end{align}
where the term $\mathbf{x}_k-\mathbf{H}_k\bm{\Phi}_k\mathbf{t}_{k-1}^s$ represents the estimated error given the previous one-step latent variable $\mathbf{t}_{k-1}^s$.

Based on the established PPFA model, the derivative of latent variables which demonstrates the system dynamic changes can be easily obtained.
\begin{equation}\label{eq:daltat}
    \bm{\delta}\left(\mathbf{t}_k^s\right)=\mathbf{t}_k^s-\mathbf{t}_{k-1}^s
\end{equation}
Besides, during the execution of E-step, the term $\mathbf{D}=\mathbb{E}\left[\bm{\delta}\left(\mathbf{t}_k^s\right)\bm{\delta}\left(\mathbf{t}_k^s\right)^\top\right]$ can be obtained as follows. 
\begin{equation}\label{eq:D}
    \mathbf{D}=\mathbb{E}\left[\mathbf{t}_k^s{\mathbf{t}_k^s}^\top\right]-2\mathbb{E}\left[\mathbf{t}_k^s{\mathbf{t}_{k-1}^s}^\top\right]+\mathbb{E}\left[\mathbf{t}_{k-1}^s{\mathbf{t}_{k-1}^s}^\top\right]
\end{equation}
Motivated by the nonstationary PSFA \citep{scott2020holistic}, a monitoring index named Dynamic Index (DI) is developed as a supplement of $\mathrm{T}^2$ and SPE to reflect the dynamics of the process.
\begin{equation}\label{eq:DI}
    \mathrm{DI}= \bm{\delta}\left(\mathbf{t}_k^s\right)^\top\mathbf{D}^{-1} \bm{\delta}\left(\mathbf{t}_k^s\right)
\end{equation}
The control limits of $\mathrm{T}^2$, SPE and DI, denoted as $\psi_{\mathrm{T}^2}$, $\psi_{\mathrm{SPE}}$ and $\psi_{\mathrm{DI}}$ respectively, are determined with the kernel density estimation (KDE), which is an effective non-parametric tool \citep{botev2010kernel, martin1996non}. Using KDE, the probability density function (pdf) of monitoring statistics is defined as
\begin{align}\label{KDE}
   f(\nu)=\frac{1}{nh}\sum_{i=1}^nK(\frac{\nu-\nu_i}{h})
\end{align}
where $h$ is the bandwidth parameter, $\nu$ represents the monitoring statistics, and $K(\cdot)$ is a kernel function selected as Gaussian in this paper.

With the estimated pdf, the cumulative density function is defined as
\begin{align}\label{CDF}
   \mathrm{P}(\nu<\psi)=\int_{- \infty}^{\psi}f(\nu)d\nu
\end{align}
where the control limit $\psi$ is obtained by setting a confidence level $\alpha$, for example, 95\%.
\subsection{Proposed Monitoring Framework}
The proposed PPFA based process monitoring method consists of two interrelated parts, namely offline modeling and online monitoring. Figure \ref{pic:Recon} \citep{ruiz2015statistical} illustrates the schematic diagram of the whole monitoring framework. For the offline modeling procedure, the first step is to normalize the normal training data. In this work, the whitening procedure is applied 
\begin{equation}\label{eq:normlization}
    \mathbf{x}_\text{norm}=\bm{\Lambda}^{-1/2}\mathbf{U}^\top\mathbf{x}
\end{equation}
where $\bm{\Lambda}$ represents the singular values of the covariance matrix $\bm{\Omega}=\langle \mathbf{x}_k \mathbf{x}_k^\top \rangle$, and $\mathbf{U}$ is an orthogonal matrix composed of eigenvectors of $\mathbf{\Omega}\mathbf{\Omega}^\top$. It is noted that the matrices $\bm{\Lambda}$ and $\mathbf{U}$ are calculated by applying  SVD to  $\bm{\Omega}$, where $\bm{\Omega}=\mathbf{U}\bm{\Lambda}\mathbf{U}^\top$.

After the normalization step, EM algorithm is applied to estimate the parameter set $\bm{\Theta}$. It is noted that in the E-step, as shown in Eq. \eqref{eq:kal1}, original data as well as parameters are augmented to simplify the Kalman smoothing process. Moreover, when executing the M-step, GA algorithm is applied to provide support for estimating $\mathbf{B}_j$. When the E-and-M-step loop iterates to the maximum number of iterations or convergence is acquired, the final parameter set $\bm{\Theta}$ is obtained. Thereafter, according to Eqs. \eqref{eq:T2} to \eqref{CDF}, the control limits $\psi_{\mathrm{T}^2}$, $\psi_{\mathrm{SPE}}$ and $\psi_{\mathrm{DI}}$, along with the monitoring statistics $\mathrm{T}^2$ , $\mathrm{SPE}$ and $\mathrm{DI}$ can be calculated.

When it comes to the online monitoring process, the newly observed sample is first scaled with Eq. \eqref{eq:normlization}. Then applying the established PPFA model to calculate the augmented latent variable $\mathbf{t}_k^s$ conditioned on the newly measurement. Further, compute the monitoring indices $\mathrm{T}^2$, $\mathrm{SPE}$ and $\mathrm{DI}$  according to Eqs. \eqref{eq:T2} to \eqref{eq:DI}. Finally, monitor if $\mathrm{T}_\text{new}^2$, $\mathrm{SPE}_\text{new}$ or $\mathrm{DI}_\text{new}$ exceeds its corresponding control limit.
\begin{itemize}
    \item A dynamic process-relevant fault or a operating condition shift is detected with $(1-\alpha)\times 100\%$ confidence level if $\mathrm{T}_\text{new}^2>\psi_{\mathrm{T}^2}$ or $\mathrm{DI}_\text{new}>\psi_{\mathrm{DI}}$.
    \item If $\mathrm{SPE}_\text{new}>\psi_{\mathrm{SPE}}$, a fault that breaks the correlation of the established model is declared. 
    \item The newly observed data is regarded as normal if $\mathrm{T}_\text{new}^2<\psi_{\mathrm{T}^2}$, $\mathrm{SPE}_\text{new}<\psi_{\mathrm{SPE}}$ and $\mathrm{DI}_\text{new}<\psi_{\mathrm{DI}}$.
\end{itemize} 
\section{Experiments and Discussion}
Two industrial processes, a three-phase flow facility and a medium speed coal mill, are employed in this section to illustrate the performance of the proposed monitoring scheme. 

\begin{figure*}[h!]
    \centering
    \includegraphics[width=6in]{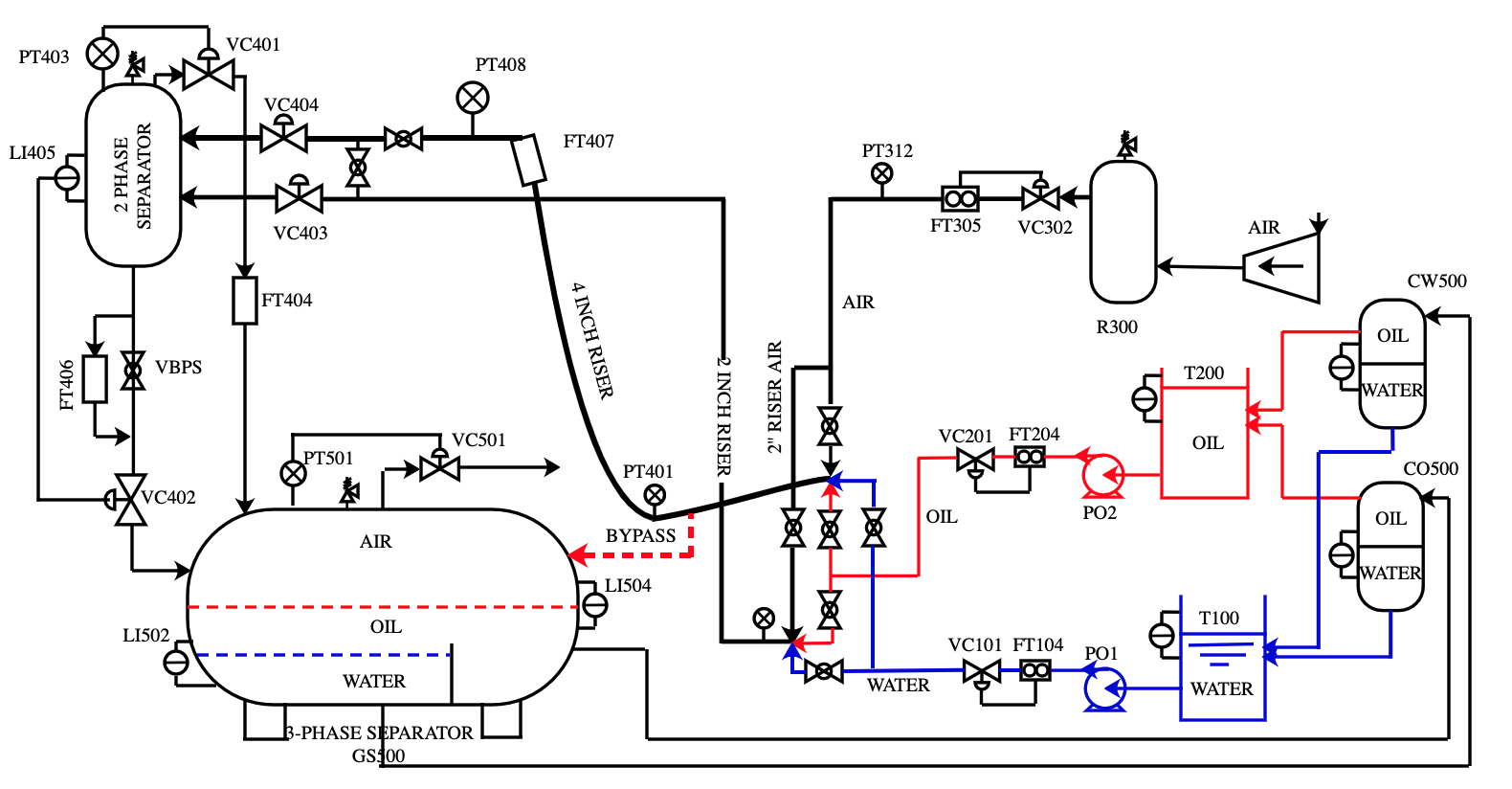}
    \caption{Sketch of three-phase flow facility.}
    \label{pic:3Phase}
\end{figure*}
\subsection{The Three-Phase Flow Facility}
\subsubsection{Process Description}
The three-phase flow facility is a pressured system suggested by Granfield University, and it is designed for feeding a controlled and measured flow rate of water, oil and air to a pressurized system \citep{ruiz2015statistical}. Recently, this complex system has received considerable attention and been successfully applied as a useful industrial process to evaluate the effectiveness of process monitoring methods \citep{zhang2017stationarity,yang2018performance}. As illustrated in the process diagram in Figure \ref{pic:3Phase}, the system is mainly composed of a two-phase separator and a three-phase separator which are connected with pipelines. It can be used to provide different products, among them are single-phase water, air and oil, or their mixtures. 

A total of 24 process variables are included in the three-phase flow facility, and at a sample rate of one second, 16 sets of data are collected, the first three of which represent the normal operation status, and the remaining sets denote faulty cases of different operating conditions. For the 24 variables, the first 23 are included in all data sets, while the last is only used in fault case 6. To generate representative normal data sets, four different set points of air flow rates and five different water flow rate set points are adopted and combined into 20 different operating conditions. Set points of air flowrate and water flowrate are listed in TABLE \ref{tab:set_points}. More details of this process are provided in Ref. \citep{ruiz2015statistical}.
\begin{table}[t]
  \centering
  \caption{Set working conditions of air flow rates $\left(\mathrm{m^3/s}\right)$ and water flow rates $\left(\mathrm{kg/s}\right)$}
    \begin{tabular}{cc}
    \toprule
    Air flowrate & 0.0208 \ \    0.0278 \ \     0.0347 \ \      0.0417 \\
    \midrule
    Water flowrate  & 0.5\ \  \ \ \ \ 1 \ \ \ \ \ \ 2 \ \  \ \ \ \ 3.5\ \  \ \ \ \ 6 \\
    \bottomrule
    \end{tabular}%
  \label{tab:set_points}%
\end{table}%
\begin{table}[t]
  \centering
  \caption{Description of data samples for modeling and monitoring}
   \begin{tabular}{lcccl}
    \toprule
        \multirow{2}{*}{Index} & \multicolumn{1}{l}{Water } & \multicolumn{1}{l}{Air} & \multicolumn{1}{l}{No. } & \multirow{2}{*}{Description} \\
          & \multicolumn{1}{l}{flowrate} & \multicolumn{1}{l}{flowrate} &  \multicolumn{1}{l}{samples}     &  \\
    \midrule
    Training data & \multirow{3}[2]{*}{2} & \multirow{3}[2]{*}{0.0417} & 3200  & Normal \\
    Test case 1 &       &       & 4467  & Airline blockage \\
    Test case 2 &       &       & 3851  & Open direct bypass \\
    \bottomrule
    \end{tabular}%
  \label{tab:data_for_model}%
\end{table}%

\subsubsection{Monitoring Results and Discussion}
In this case study, 3200 normal samples under the combination of 0.047$\mathrm{m^3/s}$ (air flow rates) and 2$\mathrm{kg/s}$ (water flow rates) are selected as the training set. In addition, two typical fault cases, as listed in TABLE \ref{tab:data_for_model}, are selected to verify the performance of the proposed method. To further illustrate the superiority of PPFA model, DiPCA \citep{dong2018novel} and PFA \citep{richthofer2015predictable} are utilized to make comparisons.

Through cross validation, parameters of different models are selected: for DiPCA, the dynamic order $s$ is set to be 3, and the number of dynamic latent variables $r=10$. For PFA, the dimension of latent variables is 10, and the time lags is chosen as 5. In the proposed PPFA, the dynamic order $s$ is set to be 2, and the dimension of $\mathbf{t}_k$ is 10. Besides, the confidence level of all methods is chosen as $99\%$.

\begin{figure}[h!]
    \centering
    \includegraphics[width=4.5in]{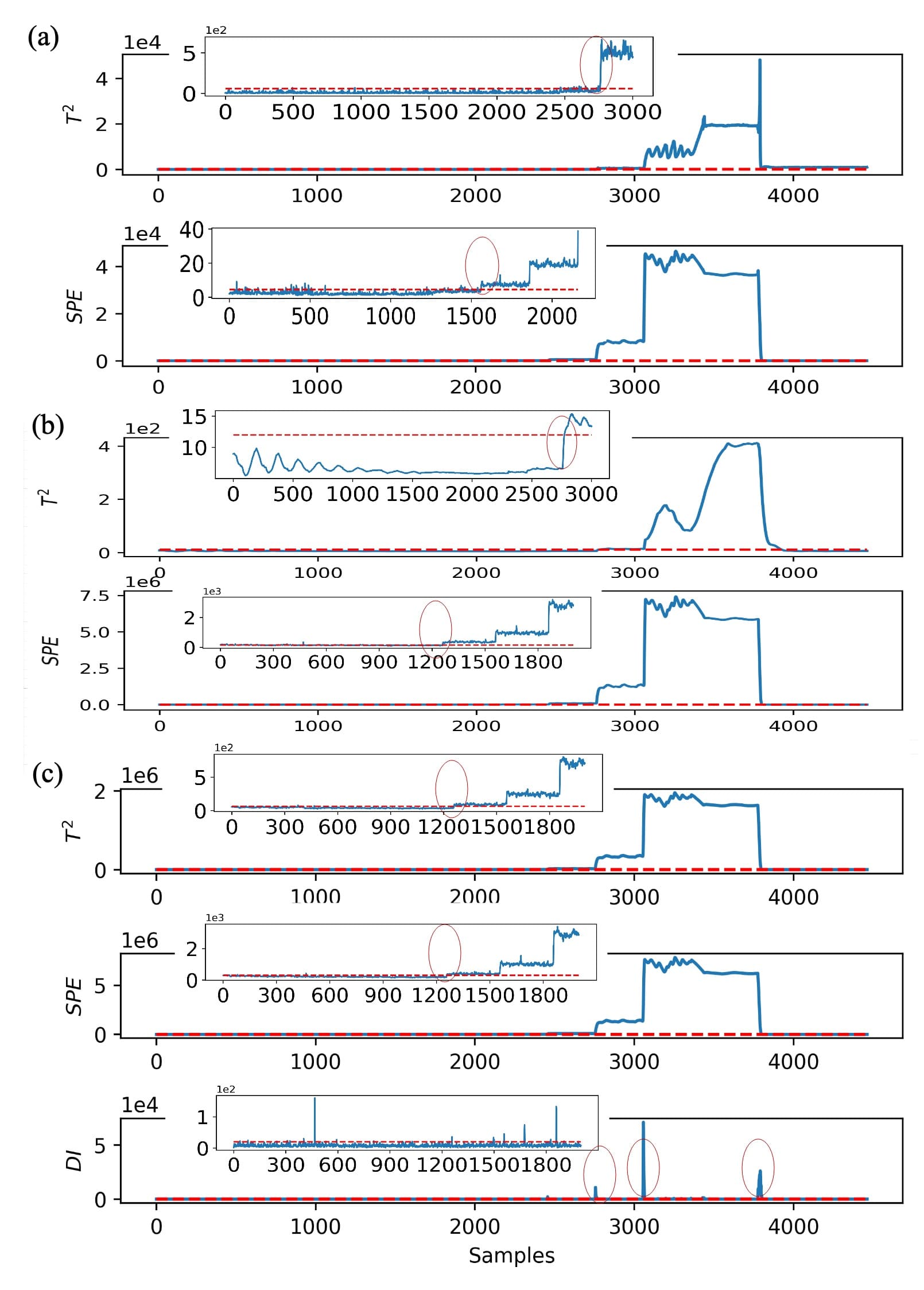}
    \caption{Monitoring results of test case 1 with (a)DiPCA, (b)PFA, and (c)PPFA.}
    \label{pic:Threephase_case1}
\end{figure}

\begin{figure}[h!]
    \centering
    \includegraphics[width=4.5in]{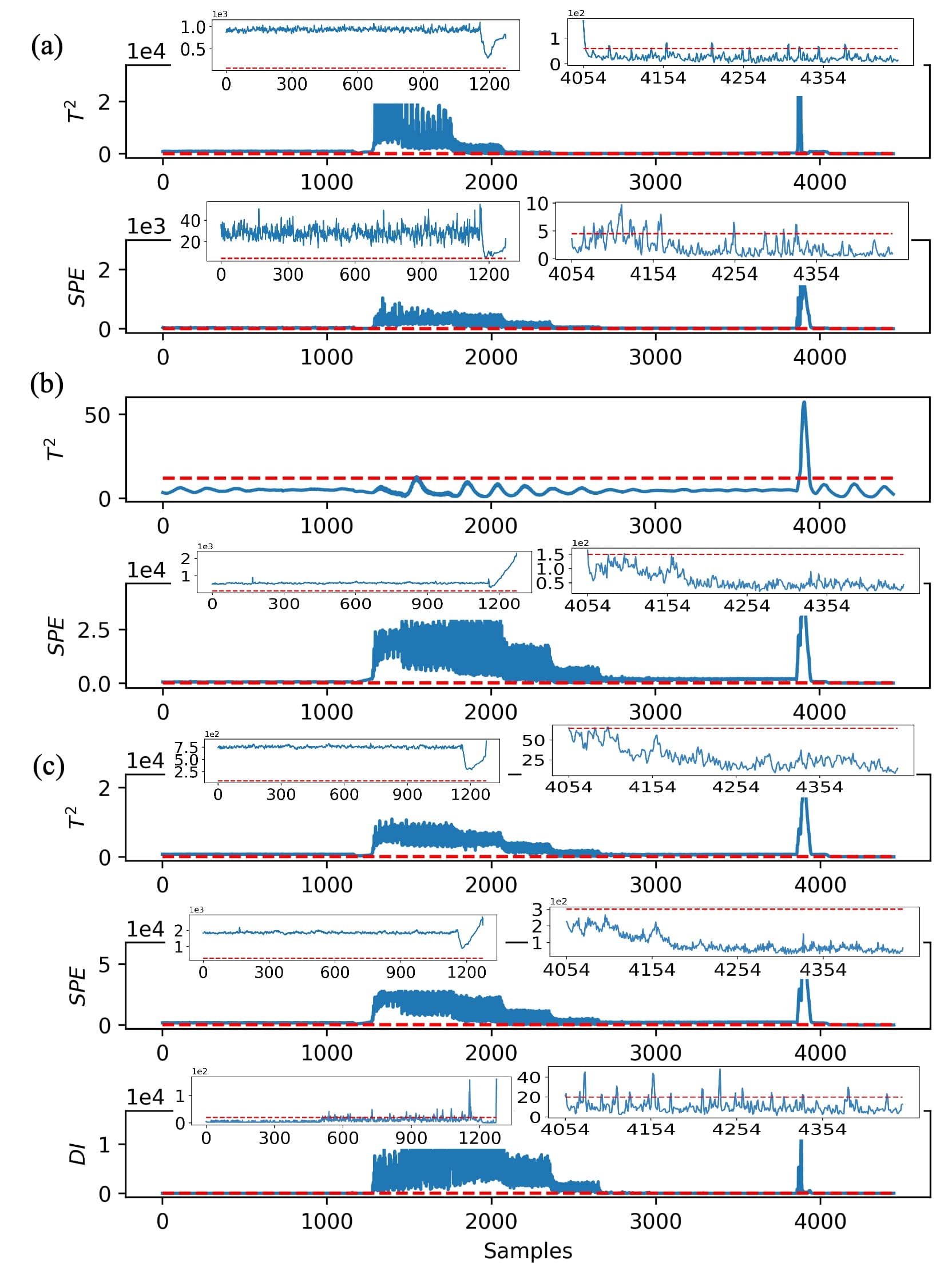}
    \caption{Monitoring results of test case 2 with (a)DiPCA, (b)PFA, and (c)PPFA.}
    \label{pic:Threephase_case2}
\end{figure}

Figure \ref{pic:Threephase_case1} and Figure \ref{pic:Threephase_case2} illustrate the monitoring results of test case 1 and test case 2 for different models. For test case 1, a total of 4467 measurements are collected. The fault of airline blockage is introduced continuously, and it starts from the $657^\mathrm{th}$ sample and ends at the $3777^\mathrm{th}$ sample. At the very beginning, the magnitude of fault is small, and the fault becomes more and more significant and reaches the maximum deviation between $3067^\mathrm{th}$ and $3777^\mathrm{th}$ sample. To ensure the security, the fault is removed from $3777^\mathrm{th}$ sample, and since then the system returns to the normal condition. From Figure \ref{pic:Threephase_case1}(c), it is observed that both $\mathrm{T}^2$ and $\mathrm{SPE}$ of PPFA start to exceed the control limit at about $1267^\mathrm{th}$ sample, then increase as the magnitude of the fault increases, and finally decrease after the fault is removed. For the new proposed index $\mathrm{DI}$, there are three peaks, reflecting the actual significant dynamic variations in the system near the $2760^\mathrm{th}$, $3067^\mathrm{th}$ and $3777^\mathrm{th}$ samples. While for DiPCA in (a) part, $\mathrm{T}^2$ stays inside the normal range from the very beginning to the $2760^\mathrm{th}$ sample, and its $\mathrm{SPE}$ goes beyond its control limit from $1573^\mathrm{th}$ sample. It is apparent that DiPCA experiences a large delay of defecting faults compared with PPFA. In addition, though $\mathrm{SPE}$ of PFA in Figure \ref{pic:Threephase_case1}(b) surpasses the limit since $1267^\mathrm{th}$ sample, $\mathrm{T}^2$ tends to reach the outer space of the control limit from $2760^\mathrm{th}$ sample, which is also later than PPFA.

In fault case 2, a fault of leakage is simulated by gradually opening the valve of the $4"$ bypass line from the $851^\mathrm{st}$ sample. The fault ends at $3851^\mathrm{st}$ and after that the process returns to a normal state. From the monitoring results in Figure \ref{pic:Threephase_case2}(a), it can be seen that both $\mathrm{T}^2$ and $\mathrm{SPE}$ of DiPCA exceed the corresponding limit from the very beginning and return to normal condition from around the $4054^\mathrm{th}$ sample. In Figure \ref{pic:Threephase_case2}(b), $\mathrm{T}^2$ is generally under the control limit and only enters the abnormal area at around the $3900^\mathrm{th}$ sample, leading to a poor monitoring performance. $\mathrm{SPE}$ stays outside of the normal region from the first sample and returns back to normal area from the $4054^\mathrm{th}$ sample. Obviously, both DiPCA and PFA provide limited information when monitoring fault case 2, and the model performance of PFA is worse. Figure \ref{pic:Threephase_case2}(c) presents the monitoring performance of PPFA, it is observed that $\mathrm{T}^2$ and $\mathrm{SPE}$ enter the abnormal region before the fault is introduced, which is similar to DiPCA and PFA. However, the novel designed dynamic index $\mathrm{DI}$ stays inside the region of normal condition before introducing the fault. It indicates that before 851s, the process is under good control, and the operating condition is different from the set point of the training set. At about the $1276^\mathrm{th}$ sample, $\mathrm{DI}$ starts to exceed the control limit, which aligns with the actual situation where the fault becomes severe. That is, the dynamic condition is disrupted from 1276s and the fault alarm of $\mathrm{DI}$ is worthy of attention.

Therefore, based on the above discussions, in PPFA, not only the fault but the dynamic variations of the system are involved, which shows a great superiority over DiPCA and PFA.

\subsection{The Medium Speed Coal Mill}
\subsubsection{Brief Description}
\begin{figure}[h!]
    \centering
    \includegraphics[width=4.5in]{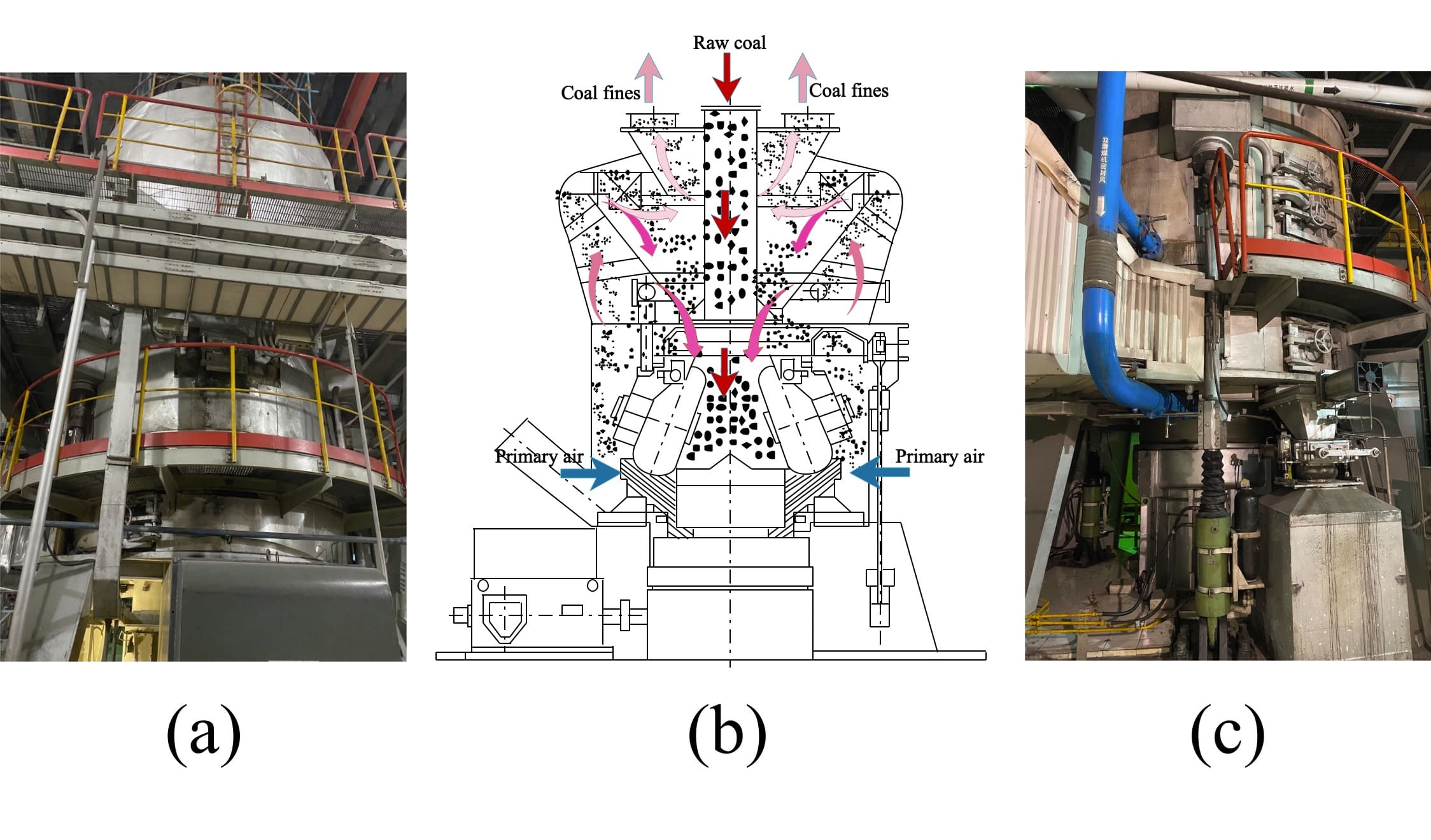}
    \caption{Schematic diagram of the medium coal mill (b)sketch, (a)(c)actual picture. }
    \label{pic:coal_fig}
\end{figure}

To further testify the effectiveness of PPFA, two different practical fault cases happened in a ZGM-113N medium speed coal mill are employed in our work. Figure \ref{pic:coal_fig} depicts the schematic diagram and actual pictures of the studied coal mill. For a general medium speed coal mill, the raw materials are transported by the coal feeder into the internal space through the inlet pipe line. The raw coal falls on a grinding table rotating at a constant speed, which is further ground to coal fines. The powder that meets a certain fineness requirement is blown into the furnace by the hot primary air for combustion. The unqualified particles fall back to the coal mill under the action of gravity and inertia, and continue to be ground. As the essential auxiliary equipment, the operating status of the coal mill has an important impact on the safety and economy of the coal-fired power plant \citep{fan2021novel}. There are mainly four typical failures during the operation process, namely intrusion of foreign materials, choking, shortage of coal and fire or explosion in the mill \citep{agrawal2016intelligent}. Once a fault occurs, it is crucial to detect it as soon as possible, otherwise it may deteriorate to an irreversible damage. To sum up, discovering and maintaining different faults in time does help to ensure the safe operation of the unit and avoid economic losses.

\subsubsection{Monitoring Results and Discussion}
\begin{figure}[h!]
    \centering
    \includegraphics[width=4.5in]{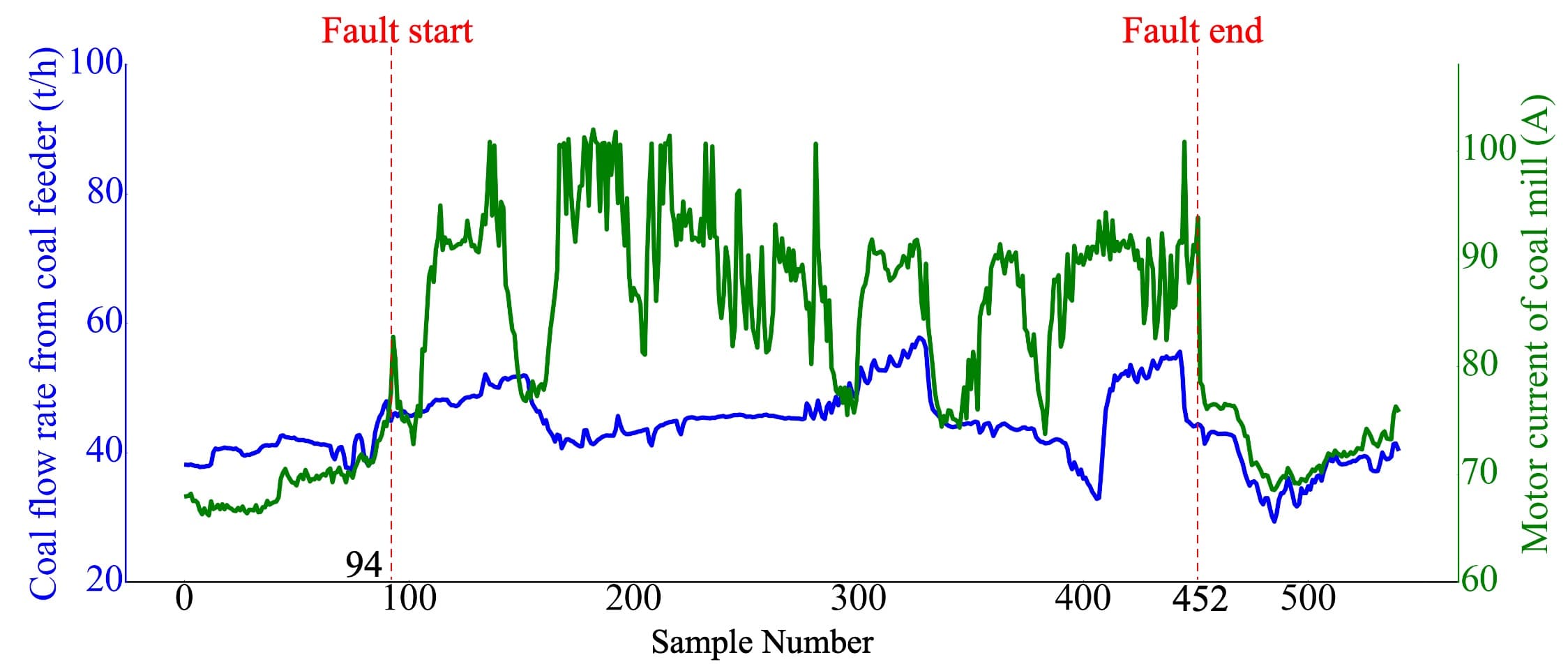}
    \caption{Historical trends of two sensors of fault case 1. }
    \label{pic:coal_hist1}
\end{figure}

In this process, two practical fault cases, mechanical failure and coal blockage, are studied. As listed in Table \ref{tab:Coal_variables}, 15 different variables are selected to build models according to the prior knowledge of coal mills \citep{fan2021novel,cortinovis2013nonlinear}. A total of normal 2385 samples are selected as the training set, and the sampling interval of the data samples is 20s. The parameters of different models are selected via cross validation: for DiPCA, $r=10$,  $s=4$; for PFA, $r=8$,  $s=5$; and for PPFA, $r=8$,  $s=3$.
\begin{table}[t]
  \centering
  \caption{Selected variables for constructing process monitoring model.}
    \begin{tabular}{llll}
    \toprule
    No.   & Variables & Description & Unit \\
    \midrule
    M1    & $\mathrm{N}_\text{unit}$     & Unit load & MW \\
    M2    & $\mathrm{M}_\text{coal}$     & Coal flow rate transported by coal feeder & t/h \\
    M3    & $\mathrm{W}_\text{air}$      & Flow rate of primary air  & t/h \\
    M4    & $\mathrm{T}_\text{air}$     & Temperature of primary air  & °C \\
    M5    & $\mathrm{P}_\text{air}$      & Pressure of primary air & kPa \\
    M6    & $\mathrm{I}_\text{mill}$     & Current of the mill's motor  & A \\
    M7    & $\mathrm{I}_\text{feed}$     & Current of coal feeder's motor  & A \\
    M8    & $\Delta\mathrm{P}_\text{seal}$     & Difference between the pressure of seal air  & kPa \\
    \multicolumn{1}{r}{} & \multicolumn{1}{r}{} & and primary air & \multicolumn{1}{r}{} \\
    M9    & $\mathrm{T}_\text{coal-air}$     & Outlet temperature of the mixture of& °C \\
     \multicolumn{1}{r}{} & \multicolumn{1}{r}{} &  coal and air  & \multicolumn{1}{r}{} \\
    M10   & $\mathrm{N}_\text{coal-air}$     & Coal-air mixture outlet pressure & kPa \\
    M11   & $\Delta\mathrm{P}_\text{in-out}$     & Difference between the pressure of  & kPa \\
    \multicolumn{1}{r}{} & \multicolumn{1}{r}{} & inlet and  outlet of coal equipment & \multicolumn{1}{r}{} \\
    M12   & $\mathrm{T}_\text{oil}$     & Coal mill's lubricating oil temperature & °C \\
    M13   & $\mathrm{T}_\text{tile}$     & Coal mill's thrust tile temperature & °C \\
    M14   &  $\mathrm{T}_\text{bearing}$    & Bearing temperature of the motor & °C \\
    M15   & $\mathrm{T}_\text{stator}$     & Stator winding temperature of the motor & °C \\
    \bottomrule
    \end{tabular}%
  \label{tab:Coal_variables}%
\end{table}%
For the fault of mechanical failure, it lasts for two hours from the $94^\mathrm{th}$ sample to the $452^\mathrm{th}$ sample. Figure \ref{pic:coal_hist1} shows the historical trends of the coal mill current and the coal flow rate. It is observed that the current fluctuates in a larger range during the fault process compared with the normal state. After maintaining the fault, some foreign matters with considerable size were found inside the mill.

\begin{figure}[t]
    \centering
    \includegraphics[width=4.5in]{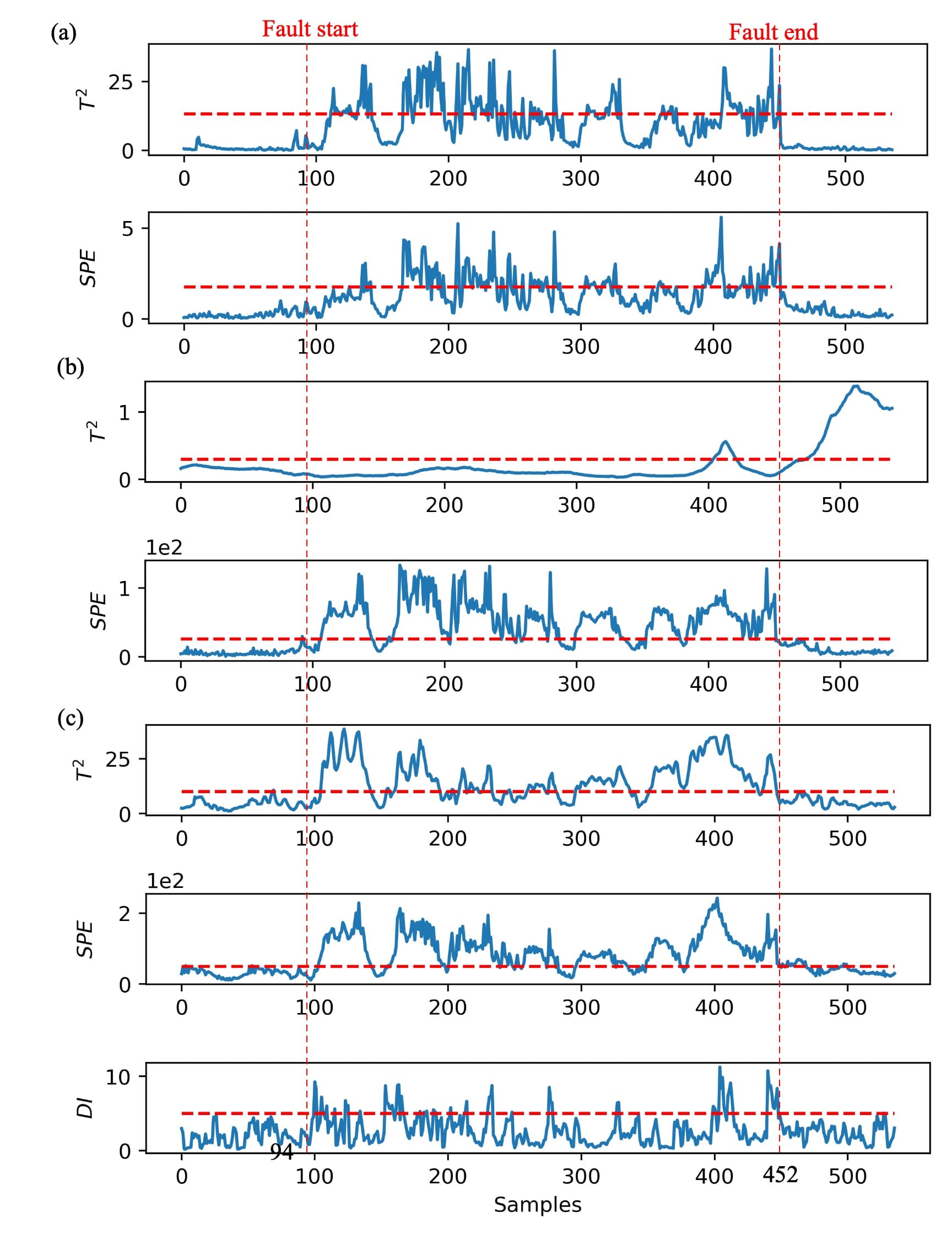}
    \caption{Monitoring results of fault case 1 with (a)DiPCA, (b)PFA, and (c)PPFA. }
    \label{pic:coal_result1}
\end{figure}

\begin{figure}[h!]
    \centering
    \includegraphics[width=6in]{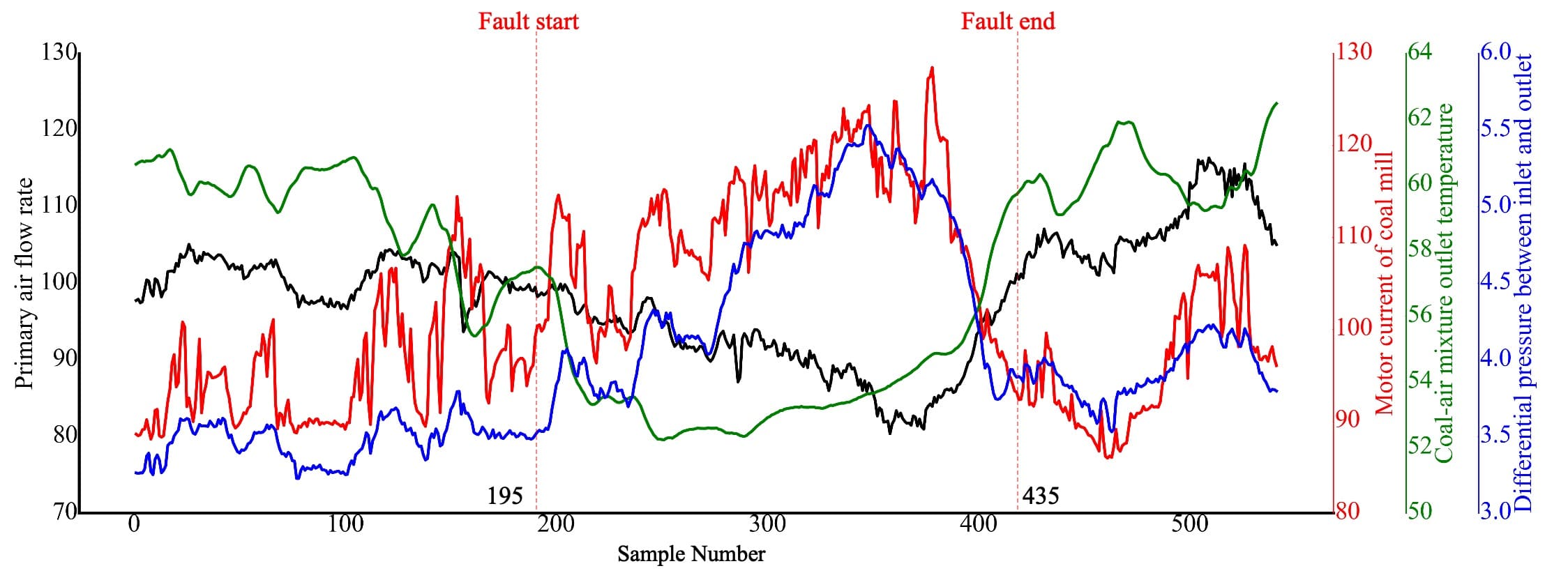}
    \caption{Historical trends of relevant sensors of fault case 2.}
    \label{pic:coal_hist2}
\end{figure}

The monitoring results of DiPCA, PFA and PPFA are presented in Figure \ref{pic:coal_result1}, in which the red dot line denotes the control limit with a 99\% confidence level. For DiPCA, as reflected in Figure \ref{pic:coal_result1}(a), both $\mathrm{T}^2$ and $\mathrm{SPE}$ begin exceeding the control limits from around the $110^\mathrm{th}$ sample, and return to normal region at the $452^\mathrm{th}$ sample. During the process when the fault occurs, a considerable portion of the two indices stay below the control limit, leading to a relatively low fault detection rate. As observed in Figure \ref{pic:coal_result1}(b), for PFA, the numerical value of $\mathrm{T}^2$ is in an  opposite trend to the actual fault, and it is lower than the red line during the occurrence of the fault, but it exceeds the threshold after the fault is removed. Though $\mathrm{SPE}$  follows the fault process in a good manner, the overall monitoring performance is unreliable. From the plots of PPFA's results illustrated in Figure \ref{pic:coal_result1}(c), it can be seen that within the two hours of failure, $\mathrm{T}^2$ and $\mathrm{SPE}$ detect the fault timely and ensure a high fault detection rate compared with DiPCA and PFA. In addition, when the coal mill current fluctuates greatly, the newly proposed index $\mathrm{DI}$ responds timely, and hence the dynamic characteristics of the system during the fault period are well reflected.
For the second fault case, as shown in Figure \ref{pic:coal_hist2}, the coal blockage starts from the $195^\mathrm{th}$ sample and ends at the $435^\mathrm{th}$ sample. When the fault exists, the mill's current and the difference between the inlet pressure and outlet pressure have an upward trend until it is discovered and handled at about the $379^\mathrm{th}$ sample. During this period, the primary air flow rate continues to decrease, and starts increasing when the fault begins recovering. The coal-air mixture temperature at the export of the mill drops significantly at the beginning of the fault, and then remains at a low value until the failure ends. During the actual operation of the unit, it takes more than one hour for the fault to be discovered, which significantly delays the maintenance of the mill and causes unnecessary economic losses.

\begin{figure}[h!]
    \centering
    \includegraphics[width=4.5in]{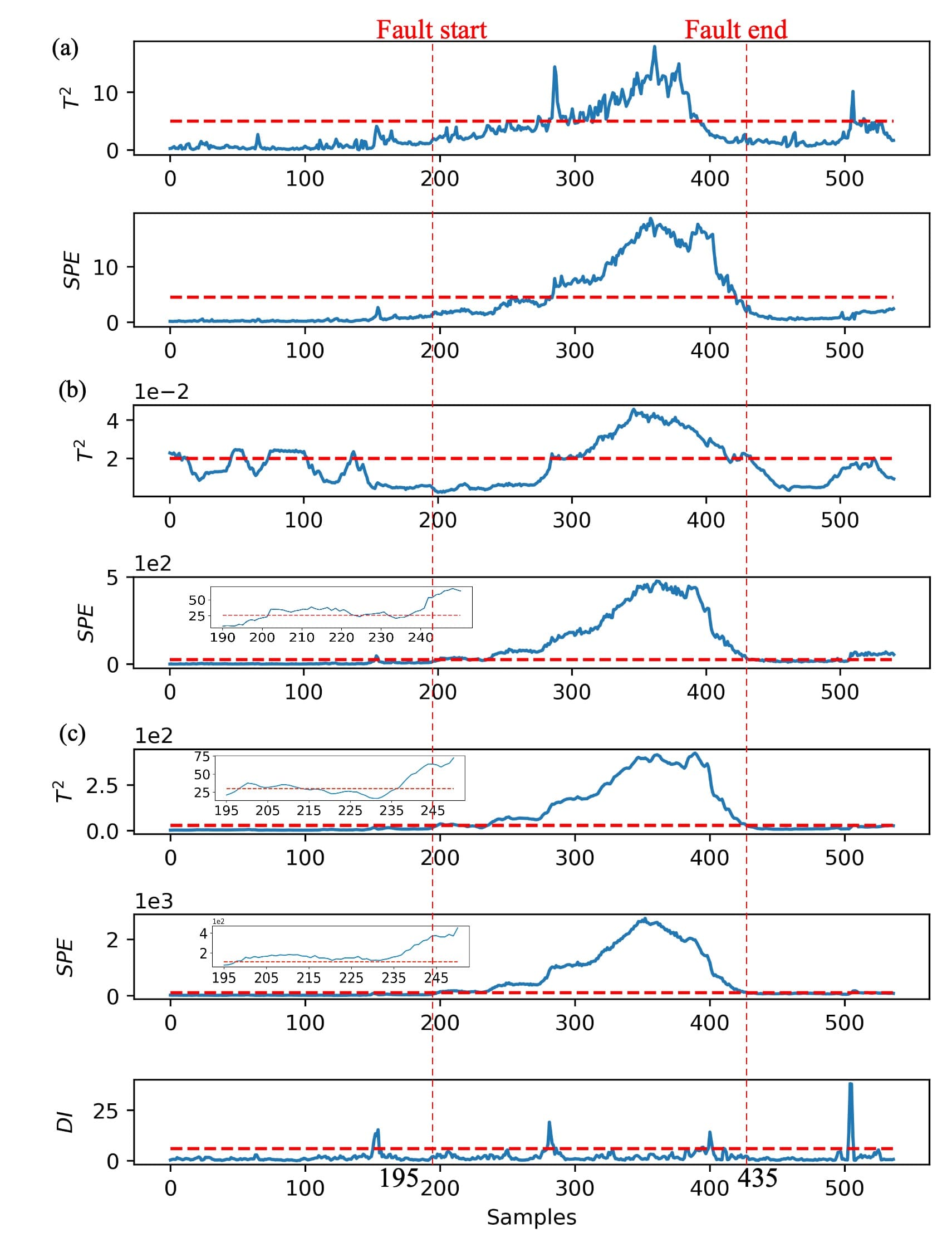}
    \caption{Monitoring results of coal blockage with (a)DiPCA, (b)PFA, and (c)PPFA. }
    \label{pic:coal_result2}
\end{figure}

Figure \ref{pic:coal_result2} presents the monitoring results for the fault of coal blockage. From Figure \ref{pic:coal_result2}(a), for the statistics $\mathrm{T}^2$ and $\mathrm{SPE}$ it can be seen that there is a delay for near 100 samples (0.56 hour) after the disturbance happens. For the results of PFA, depicted in Figure \ref{pic:coal_result2}(b), it shows similar pattern as it does for fault case 1. That is, though $\mathrm{SPE}$ retains a satisfying performance, $\mathrm{T}^2$ has a poor performance which raises a large number of false alarms and obtains a low fault detection rate. For the proposed PPFA model, it is observed in Figure \ref{pic:coal_result2}(c) that $\mathrm{T}^2$ and $\mathrm{SPE}$ successfully raise the alarm once the fault occurs, along with a great fault detection rate. Moreover, combining with the trend of the variables in Figure \ref{pic:coal_hist2}, it can be seen that $\mathrm{DI}$ has obvious over-limit response to rapid fluctuations in different stages. In conclusion, considering the monitoring performance, the proposed PPFA model has great advantages over existing methods such as DiPCA and PFA. Its advantages are mainly manifested in that this method can not only  realize fault detection quickly and accurately, but also the newly proposed index can be used as a supplement to reflect the dynamic changes of the system.

\section{Conclusion}
In this paper, a novel probabilistic predictable feature analysis method is proposed for multivariate time series monitoring. The proposed method takes measurement noise and full interpretation of dynamic characteristics into consideration. During the procedure of EM iterations, GA and Kalman filter are successfully employed to estimate parameters. In addition, a dynamic index (DI) along with $\mathrm{T}^2$ and $\mathrm{SPE}$ 
is developed and applied to monitor the dynamic operation conditions of industrial processes. Through applications on the three-phase flow facility and a medium speed coal mill, the PPFA based process monitoring method has shown its effective performance compared with existing deterministic methods. Based on the monitoring results, the PPFA algorithm is worth for further investigation. 

\appendix
\section{Proof of \textit{Lemma \ref{Lemma1}}}
\label{appendixA}

According to Eq. \eqref{PPFA_system}, we have
\begin{equation}\label{eq:app1}
    t_k^i=\beta_1^it_{k-1}^i+\cdots+\beta_s^it_{k-s}^i+e_k^i
\end{equation}
With the constraint $\mathbb{E}\left[\mathbf{t}_k\right] = \mathbf{0}$, the expectation of $t_k^i$ is
\begin{equation}\label{eq:app2}
    \mathbb{E}\left[t_k^i\right]=\beta_1^i \mathbb{E}\left[t_{k-1}^i\right]+\cdots+\beta_s^i \mathbb{E}\left[t_{k-s}^i\right]+\mathbb{E}\left[e_k^i\right]=0
\end{equation}
The variance of $t_k^i$ can be obtained with the following formula
\begin{equation}\label{eq:app3}
    \textup{var}\left(t_k^i\right)=\mathbb{E}\left[\left(t_k^i\right)^2\right]-\mathbb{E}\left[t_k^i\right]^2=\mathbb{E}\left[\left(t_k^i\right)^2\right]
\end{equation}
Since $\left(t_k^i\right)^2=\beta_1^it_k^it_{k-1}^i+\cdots+\beta_s^it_k^it_{k-s}^i+t_k^ie_k^i$, Eq. \eqref{eq:app3} is computed by
\begin{equation}\label{eq:app4}
    \mathbb{E}\left[\left(t_k^i\right)^2\right]=\beta_1^i\mathbb{E}\left[t_k^i t_{k-1}^i\right]+\cdots+\beta_s^i\mathbb{E}\left[t_k^i t_{k-s}^i\right]+\mathbb{E}\left[t_k^i e_k^i\right]
\end{equation}
where the last term can be further simplified by
\begin{equation}\label{eq:app5}
\begin{split}
    \mathbb{E}\left[t_k^i e_k^i\right]&=\mathbb{E}\left[\left(\beta_1^it_{k-1}^i+\cdots+\beta_s^it_{k-s}^i+e_k^i \right)e_k^i\right]\\
    &=\mathbb{E}\left[\left(\beta_1^it_{k-1}^i+\cdots+\beta_s^it_{k-s}^i \right)e_k^i\right]+\mathbb{E}\left[\left(e_k^i\right)^2\right]\\
    &=\mathbb{E}\left[\beta_1^it_{k-1}^i+\cdots+\beta_s^it_{k-s}^i\right]\mathbb{E}\left[e_k^i\right]+\mathbb{E}\left[\left(e_k^i\right)^2\right]\\
    &=\mathbb{E}\left[\left(e_k^i\right)^2\right]=\tau_i^2
\end{split}
\end{equation}
For ease of representation, we define
\begin{equation}\label{eq:app6}
\begin{split}
     \gamma_j^i&=\mathrm{cov}\left(t_k^i,t_{k-j}^i\right)=\mathrm{cov}\left(t_{k+j}^i,t_k^i\right)\\
     &=\mathbb{E}\left[t_k^it_{k-j}^i\right]-\mathbb{E}\left[t_k^i\right]\mathbb{E}\left[t_{k-j}^i\right]\\
     &=\mathbb{E}\left[t_k^it_{k-j}^i\right]
\end{split}
\end{equation}
Therefore, the covariance in Eq. \eqref{eq:app3} simplifies to
\begin{equation}\label{eq:app7}
    \textup{var}\left(t_k^i\right)=\mathbb{E}\left[\left(t_k^i\right)^2\right]=\beta_1^i\gamma_1^i+\cdots+\beta_s^i\gamma_s^i+\tau_i^2
\end{equation}
Since the constraint $\mathbb{E} \left[\mathbf{t}_k \mathbf{t}_k^\top\right]=\mathbf{I}_{r}$ has been given in  Eq. \eqref{PPFA_system}, we have
\begin{equation}\label{eq:app8}
    \beta_1^i\gamma_1^i+\cdots+\beta_s^i\gamma_s^i+\tau_i^2=1
\end{equation}
Therefire, the relation between $\mathbf{B}_j$ and $\mathbf{\Gamma}$ can be expressed by Eq. \eqref{eq:Lemma1}.

\section*{Acknowledgment}

This work was supported by China Scholarship Council (grant numbers 202006090212), Qinglan Project of Jiangsu Province of China, National Natural Science Foundation of China under Grant 51976031 as well as the University of Waterloo.This work has been published by IEEE Transactions on Control Systems Technology. Copyright has been transferred without notice, after which this version may no longer be accessible. 

\bibliographystyle{unsrtnat}
\bibliography{PPFA}

\end{document}